\documentclass[english,twocolumn,showpacs,preprintnumbers,amsmath,amssymb]{revtex4}
\usepackage[T1]{fontenc}
\usepackage[latin1]{inputenc}
\usepackage{graphicx}

\makeatletter

\makeatletter

\makeatletter

\makeatletter

\usepackage{dcolumn}

\usepackage{bm}

\makeatother

\makeatother

\makeatother

\usepackage{babel}
\makeatother

\begin{document}

\title{Energy bursts in fiber bundle models of composite materials}

\author{$^{1,2}$Srutarshi Pradhan}

\email{pradhan.srutarshi@ntnu.no}

\author{$^{1}$Per C. Hemmer}

\email{per.hemmer@ntnu.no}

\affiliation{$^{1}$Department of Physics, Norwegian University of Science and
Technology, N--7491 Trondheim, Norway}

\affiliation{$^{2}$SINTEF Petroleum Research, Trondheim, Norway}

\begin{abstract}
As a model of composite materials, a bundle of many fibers with stochastically
distributed breaking thresholds for the individual fibers is considered.
The bundle is loaded until complete failure to capture the failure
scenario of composite materials under external load. The fibers are
assumed to share the load equally, and to obey Hookean elasticity
right up to the breaking point. We determine the distribution of bursts
in which an amount of energy $E$ is released. The energy distribution
follows asymptotically a universal power law $E^{-5/2}$, for any
statistical distribution of fiber strengths. A similar  power law dependence
is found in some experimental acoustic emission
studies of loaded composite materials.\\

\end{abstract}


\maketitle

\section{Introduction}

During the failure process of composite materials under external load,
avalanches of different magnitudes are produced, where an avalanche
consists of simultaneous rupture of several elements. Such avalanches
cause a sudden internal stress redistribution in the material, and
are accompanied by a rapid release of mechanical energy. A useful
experimental technique to monitor the energy release is to measure
the acoustic emissions, the elastically radiated waves produced in
the bursts \cite{Petri,Ciliberto97,Scott,Fazzini,Diodati}.

Fiber bundles with statistically distributed thresholds for breakdown
of individual fibers are interesting models of failure processes in
materials. They are characterized by simple geometry and clear-cut
rules for how stress caused by a failed element is redistributed on
the intact fibers. The interest of these models lies in the possibility
of obtaining exact results, thereby providing inspiration and reference
systems for studies of more complicated materials. (For reviews, see
\cite{Herrmann,Chakrabarti,Sornette,Sahimi,Bhattacharyya}). The
statistical distribution of the \textit{size} of avalanches in fiber
bundles is well studied \cite{HH,PHH05,HHP,Kun06}, but the distribution
of the burst \textit{energies} is not. In this article we therefore
determine the statistics of the energies released in fiber bundle
avalanches.

We study equal-load-sharing models, in which the load previously carried
by a failed fiber is shared equally by all the remaining intact fibers
in the bundle \cite{Peirce,Daniels,Smith,Phoenix}. We consider a
bundle consisting of a large number $N$ of elastic fibers, clamped
at both ends. The fibers obey Hooke's law, such that the energy stored
in a single fiber at elongation $x$ equals $\frac{1}{2}x^{2}$, where
we for simplicity have set the elasticity constant equal to unity.
Each fiber $i$ is associated with a breakdown threshold $x_{i}$
for its elongation. When the length exceeds $x_{i}$ the fiber breaks
immediately, and does not contribute to the strength of the bundle
thereafter. The individual tresholds $x_{i}$ are assumed to be independent
random variables with the same cumulative distribution function $P(x)$
and a corresponding density function $p(x)$: \begin{equation}
{\rm Prob}(x_{i}<x)=P(x)=\int_{0}^{x}p(y)\; dy.\end{equation}

\begin{center}
\includegraphics[width=6cm,height=6cm]{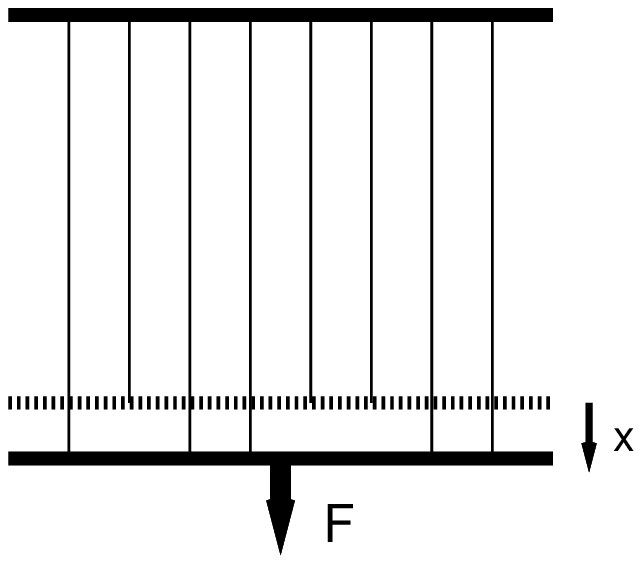}
\par\end{center}

\noindent {\small FIG.\ 1. The fiber bundle model. }\\
{\small \par}

At an elongation $x$ the total force on the bundle is $x$ times
the number of intact fibers. The average, or macroscopic, force is
given by the expectation value of this, \begin{equation}
\langle F\rangle=N\; x\;[1-P(x)].\end{equation}
 In the generic case $\langle F\rangle$ will have a single maximum
$F_{c}$, a critical load corresponding to the maximum load the bundle
can sustain before complete breakdown of the whole system. The maximum
occurs at a critical value $x_{c}$ for which $d\langle F\rangle/dx$
vanishes. Thus $x_{c}$ satisfies \begin{equation}
1-P(x_{c})-x_{c}\; p(x_{c})=0.\label{xc}\end{equation}

\section{Energy statistics}

Let us characterize a burst by the number $n$ of fibers that fail,
and by the lowest threshold value $x$ among the $n$ failed fibers.
The treshold value $x_{{\rm max}}$ of the strongest fiber in the
burst can be estimated to be \begin{equation}
x_{{\rm max}}\simeq x+\frac{n}{Np(x)},\label{xmax}\end{equation}
 since the expected number of fibers with thresholds in an interval
$\Delta x$ is given by the threshold distribution function as $N\; p(x)\;\Delta x$.
The last term in (\ref{xmax}) is of the order $1/N$, so for a very
large bundle the differences in threshold values among the failed
fibers in one burst are negligible. Hence the energy released in a
burst of size $n$ that starts with a fiber with threshold $x$ is
given with sufficient accuracy as \begin{equation}
E={\textstyle \frac{1}{2}}\; n\; x^{2}.\label{En}\end{equation}

In a statistical analysis of the burst process Hemmer and Hansen \cite{HH}
calculated the expected number of bursts of size $n$, starting at
a fiber with a threshold value in the interval $(x,x+dx)$, as 
\begin{eqnarray}
f(n,x)\; dx & = & N\frac{n^{n-1}}{n!}\;\frac{1-P(x)-xp(x)}{x}\nonumber\\
& \times & X(x)^{n}\; e^{-nX(x)}\; dx,
\end{eqnarray}
with the abbreviation \begin{equation}
X(x)=\frac{x\; p(x)}{1-P(x)}.\end{equation}
The expected number of bursts with energies less than $E$ is therefore
\begin{equation}
G(E)=\sum_{n}\int\limits _{0}^{\sqrt{2E/n}}f(n,x)\; dx,\end{equation}
 with a corresponding energy density \begin{equation}
g(E)=\frac{dG}{dE}=\sum_{n}(2En)^{-1/2}\; f(n,\sqrt{2E/n}).\end{equation}
 Explicitly, \begin{equation}
g(E)=N\sum_{n}g_{n}(E),\label{g}\end{equation}
 with 
\begin{eqnarray}
g_{n}(E)& = & \frac{n^{n-1}}{2E\; n!}(1-P(s)-sp(s))\nonumber\\
& \times & \left[\frac{sp(s)}{1-P(s)}\exp\left(-\frac{sp(s)}{1-P(s)}\right)\right]^{n}.\label{gn}\end{eqnarray}
Here 
\begin{equation}
s\equiv\sqrt{2E/n}.\label{s}
\end{equation}

With a critical threshold value $x_{c}$, it follows from (\ref{En})
that a burst energy $E$ can only be obtained if $n$ is sufficiently
large, \begin{equation}
n\geq2E/x_{c}^{2}.\end{equation}
 Thus the sum over $n$ starts with 
\begin{equation}
n=1+[2E/x_{c}^{2}],
\end{equation}
 here $[a]$ denotes the integer part of $a$.

We discuss now both the high-energy and the low-energy behavior of
the energy density $g(E)$.

\subsection{High energy asymptotics}

Bursts with high energies correspond to bursts in which many fibers
rupture. In this range we may use Stirling's approximation for the
factorial $n!$, replace $1+[2E/x_{c}^{2}]$ by $2E/x_{c}^{2}$, and
replace the summation over $n$ by an integration. Thus

\begin{eqnarray}
g(E) & \simeq & \frac{N}{2E^{3/2}\pi^{1/2}}\int\limits _{2E/x_{c}^{2}}^{\infty}\frac{e^{n}}{n^{3/2}}\;(1-P(s)-sp(s))\nonumber \\
 & \times & \left[\frac{sp(s)}{1-P(s)}\exp\left(-\frac{sp(s)}{1-P(s)}\right)\right]^{n}\; dn,\label{integral1}\end{eqnarray}
 where $s$ is the abbreviation (\ref{s}). By changing integration
variable from $n$ to $s$ we obtain {\begin{eqnarray}
g(E) & \simeq & \frac{N}{2E^{3/2}\pi^{1/2}}\int\limits _{0}^{x_{c}}(1-P(s)-sp(s))\nonumber\\ & \times & \left[\frac{sp(s)}{1-P(s)}\exp\left(1-\frac{sp(s)}{1-P(s)}\right)\right]^{n}\; ds\nonumber \\
 & = & \frac{N}{2E^{3/2}\pi^{1/2}}\int\limits _{0}^{x_{c}}(1-P(s)-sp(s))e^{-Eh(s)}\; ds,\label{integral}\end{eqnarray}
}  with \begin{equation}
h(s)\equiv\left[-\frac{1-P(s)-sp(s)}{1-P(s)}+\ln\frac{1-P(s)}{sp(s)}\right]\frac{2}{s^{2}}.\label{h}\end{equation}

For large $E$ the integral (\ref{integral}) is dominated by the
integration range near the minimum of $h(s)$. At the upper limit
$s=x_{c}$ we have $h(x_{c})=0$, since $1-P(x_{c})=x_{c}p(x_{c})$,
Eq.(\ref{xc}). This is also a minimum of $h(s)$. To see that, note
that with $y\equiv1-sp(s)/(1-P(s))$, the bracket in (\ref{h}) is
of the form \begin{equation}
-y-\ln(1-y)=y^{2}+{\cal O}(y^{3}),\end{equation}
 with a minimum at $y=0$.

In a systematic expansion about the maximum of the integrand in (\ref{integral}), at $s=x_{c}$, the first factor in the integral (\ref{integral})
vanishes linearly, 
\begin{eqnarray}
1-P(s)-sp(s) & = &(x_{c}-s)[2p(x_{c})+x_{c}p'(x_{c})]\nonumber\\
& + &  {\cal O}(x_{c}-s)^{2},\end{eqnarray}
 and, as we have seen, $h(s)$ has a quadratic minimum,
\begin{equation}
h(s)\simeq\left(\frac{2p(x_{c})+x_{c}p'(x_{c})}{x_{c}^{2}p(x_{c}}\right)^{2}\;(x_{c}-s)^{2}.\end{equation}
 Inserting these expressions into (\ref{integral}) and integrating,
we obtain the following asymptotic expression, \begin{equation}
g(E)\simeq N\;\frac{C}{E^{5/2}},\label{as}\end{equation}
 where \begin{equation}
C=\frac{x_{c}^{4}p(x_{c})^{2}}{4\pi^{1/2}\,[2p(x_{c})+x_{c}p'(x_{c})]}.\label{C}\end{equation}

\begin{center}
\includegraphics[width=6cm,height=6cm]{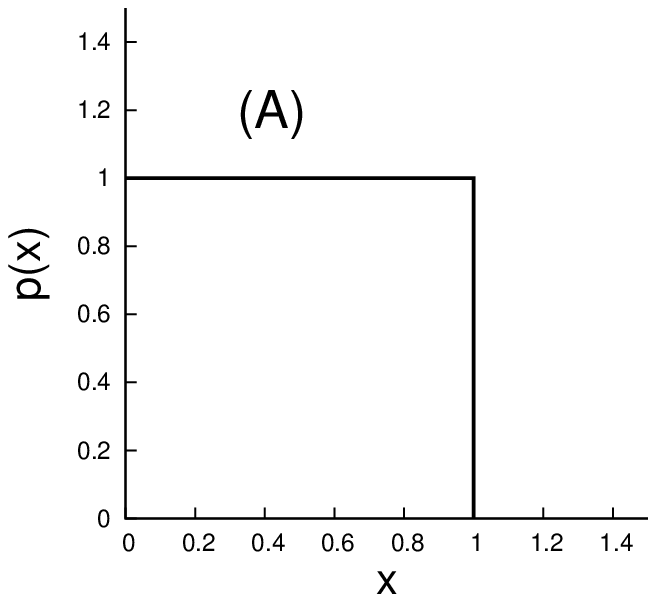}
\par\end{center}

\begin{center}
\includegraphics[width=6cm,height=6cm]{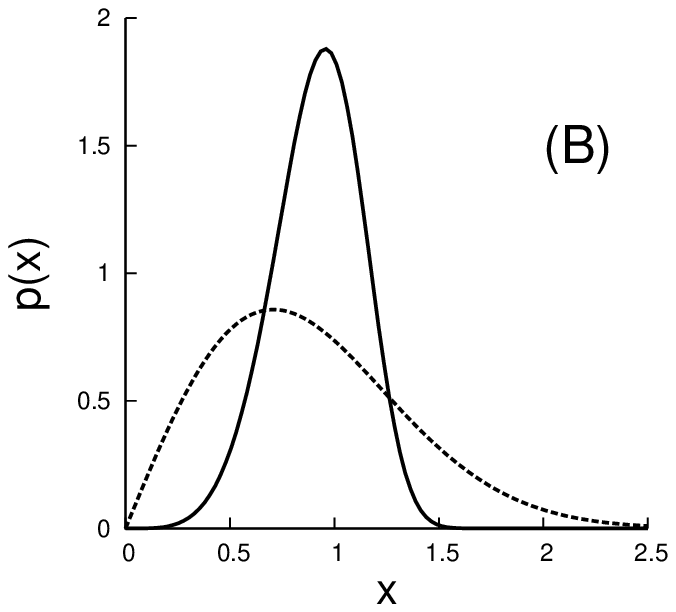}
\par\end{center}

\noindent {\small FIG.\ 2. (A) The uniform threshold distribution
(\ref{uniform}) and (B) the Weibull distribution (\ref{Weibull})
of index $2$ (dotted line) and index $5$ (solid lin}e) . \\

In Fig.\ 3 we compare the theoretical formula with simulations for
the uniform distribution, \begin{equation}
P(x)=\left\{ \begin{array}{ll}
x & \mbox{ for }0\leq x\leq1,\\
0 & \mbox{ for }x>1,\end{array}\right.\label{uniform}\end{equation}
 which corresponds to $x_{c}=\frac{1}{2}$, and $C=2^{-7}\pi^{-1/2}$,
and for the Weibull distribution with index $k=2$, \begin{equation}
P(x)=1-e^{-x^{k}}\hspace{1cm}\mbox{for }x\geq0,\label{Weibull}\end{equation}
 which corresponds to $x_{c}=2^{-1/2}$ and $C=2^{-5}(2\pi e)^{-1/2}$.

\begin{center}
\includegraphics[width=2.5in,height=2.4in]{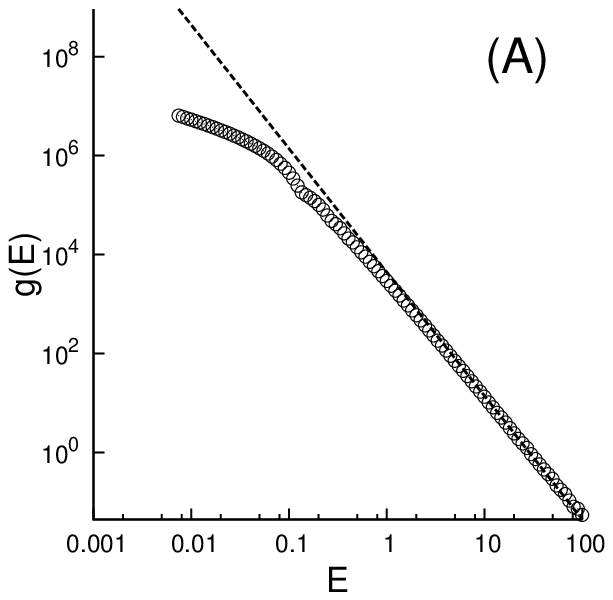} 
\par\end{center}

\begin{center}
\includegraphics[width=2.5in,height=2.4in]{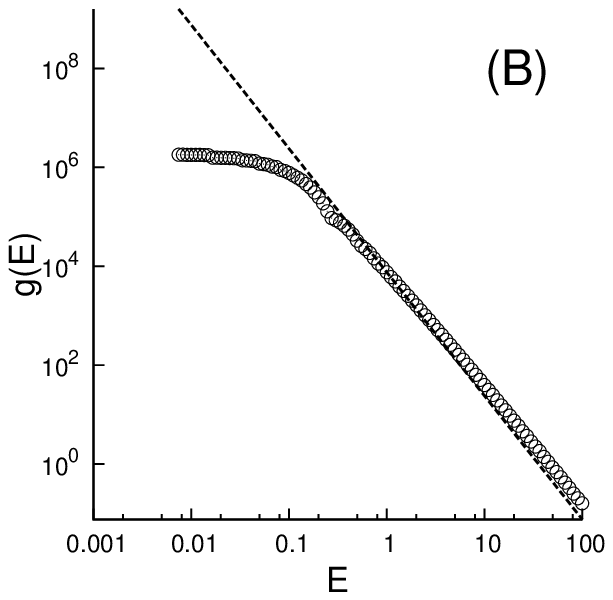} 
\par\end{center}

\noindent {\small FIG.\ 3. Simulation results for $g(E)$ characterizing
energy bursts in fiber bundles with (A) the uniform threshold distribution
(\ref{uniform}) and (B) the Weibull distribution (\ref{Weibull})
of index $2$. The graphs are based on 1000 samples with $N=10^{6}$
fibers in each bundle. Open circles represent simulation data, and
dashed lines are the theoretical results (\ref{as}-\ref{C}) for
the asymptotics. }\\
{\small \par}

The corresponding asymptotics (\ref{as}) are also exhibited in Fig.\ 3.
For both threshold distributions the agreement between the theoretical
asymptotics and the simulation results is very satisfactory. The exponent
$-5/2$ in the energy burst distribution is clearly universal. Note
that the asymptotic distribution of the burst magnitudes $n$ is governed
by the same exponent \cite{HH}.

\subsection{Low-energy behavior}

The low-energy behavior of the burst distribution is by no means universal:
$g(E)$ may diverge, vanish or stay constant as $E\rightarrow0$,
depending on the nature of the threshold distribution. In Fig.\ 4
we exhibit simulation results for the low-energy part of $g(E)$ for
the uniform distribution and the Weibull distributions of index $2$
and index $5$.

\begin{center}
\includegraphics[width=2.5in,height=2.4in]{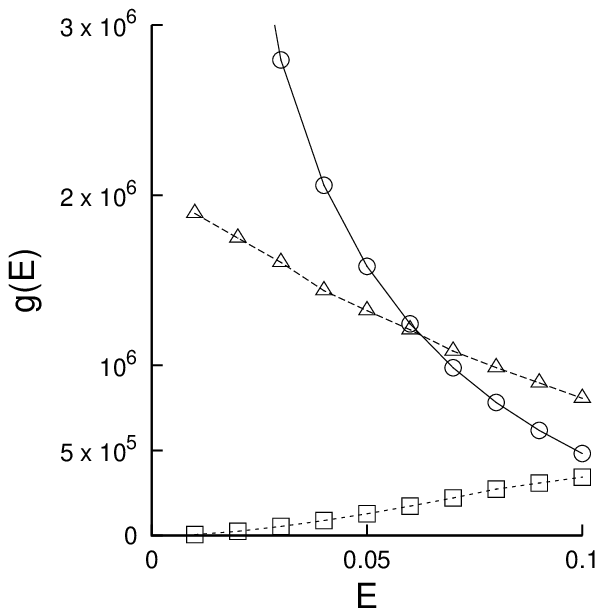} 
\par\end{center}

\noindent FIG.\ 4{\small . Simulation results for the burst distribution
$g(E)$, in the low-energy regime, for the uniform threshold distribution
(circles), the Weibull distribution with $k=2$ (triangles) and Weibull
distribution with $k=5$ (squares). The graphs are based on 1000 samples
with $N=10^{6}$ fibers in each bundle. }\\
{\small \par}

We see that $g(E)$ approaches a finite limit in the Weibull $k=2$
case, approaches zero for Weibull $k=5$ and apparently diverges in
the uniform case. All this is easily understood, since bursts with
low energy predominently correspond to single fiber bursts ($n=1$,
\textit{i.e.} $E=x^{2}/2$) and to fibers with low threshold values.
The number of bursts with energy less than $E$ therefore corresponds
to the number of bursts with $x<\sqrt{2E}$, which is close to $N\; P(\sqrt{2E})$.
This gives \begin{equation}
g(E)\simeq N\;\frac{p(\sqrt{2E})}{\sqrt{2E}}\hspace{10mm}\mbox{when }E\rightarrow0.\label{Enull}\end{equation}
 For the uniform distribution $g(E)$ should therefore diverge as
$(2E)^{-1/2}$ for $E\rightarrow0$. The simulation results in Fig.\ 4
are consistent with this divergence. For the Weibull of index 2, on
the other hand, (\ref{Enull}) gives $g(E)\rightarrow2N$ when $E\rightarrow0$,
a value in agreement with simulation results in the figure. Note that
for a Weibull distribution of index $k$, the low-energy behavior
is $g(E)\propto E^{(k-2)/2}$. Thus the Weibull with $k=2$ is a borderline
case between divergence and vanishing of the low-energy density.

The same lowest-order results can be obtained from the general expression
(\ref{g}), which also can provide more detailed low-energy expansions.
\vspace {1 cm}

\section{Summary}

In the present article we have studied the distribution of burst energies
during the failure process in fiber bundles with statistically distributed
thresholds for breakdown of individual fibers. We have derived an
exact expression for the energy density distribution $g(E)$, and
shown that for high energies the energy density obeys a power law
with exponent $-5/2$. This asymptotic behavior is universal, independent
of the threshold distribution. A similar  power law dependence is found in some 
experimental observations on acoustic emission
studies \cite{Petri,Ciliberto97} of loaded composite materials.

In contrast the low-energy behavior of $g(E)$ depends crucially on
the distribution of the breakdown thresholds in the bundle. $g(E)$
may diverge, vanish or stay constant for $E\rightarrow0$.

\begin{acknowledgments}
This work has been supported by Norwegian Research Council (NFR) through
project number $177958/V30$. S. Pradhan thanks the Formation Physics Group,
SINTEF Petroleum Research, for moral support and encouragement. 
\end{acknowledgments}


\begin{thebibliography}{10}
\bibitem{Petri}A.\ Petri, G. Paparo, A. Vespignani, A. Alippi, and
M. Constantini, Phys. Rev. Lett. \textbf{73} 3423 (1994).
\bibitem{Ciliberto97} A. Garcimartin, A. Guarino, L. Bellon and S. Ciliberto, 
Phys. Rev. Lett. \textbf{79} 3202 (1997).

\bibitem{Scott}I.\ G.\ Scott, \textit{Basic acoustic emission,
}\textit{\emph{p. 246 in}}\textit{ Nondestructive Testing Monographs and Track}
 vol \textbf{6} (Gordon and Breach Science Publishers, New York, 1991).

\bibitem{Fazzini}P.\ Fazzini, \textit{Basic Acoustic Emission},
(Gordon and Breach Science Publishers, New York, 1991).

\bibitem{Diodati} P.\ Diodati, F.\ Marchesoni, and S.\ Piazza,
Phys.\ Rev.\ Lett. \textbf{67}, 2239 (1991).

\bibitem{Herrmann} \textit{Statistical Models for the fracture of
Disordered Materials}, edited by H.\ J.\ Herrmann and S.\ Roux
(Elsvier, Amsterdam, 1990).

\bibitem{Chakrabarti} B.\ K.\ Chakrabarti and L.\ G.\ Benguigui,
\textit{Statistical Physics and Breakdown in Disordered Systems} (Oxford
University Press, Oxford, 1997).

\bibitem{Sornette} D.\ Sornette, \textit{Critical Phenomena in Natural
Sciences} (Springer-Verlag, Berlin, 2000).

\bibitem{Sahimi} M.\ Sahimi, \textit{Heterogenous Materials II:
Nonlinear and Break-down Properties} (Springer-Verlag, Berlin, 2003).Phys.
Rev. Lett. \textbf{73} 3423 (1994).

\bibitem{Bhattacharyya} \textit{Modeling Critical and Catastrophic
Phenomena in Geoscience}, edited by P.\ Bhattacharyya and B.\ K.\ Chakrabarti
(Springer-Verlag, Berlin, 2006).

\bibitem{HH} P.\ C.\ Hemmer and A.\ Hansen, ASME J.\ Appl.\ Mech.
\textbf{59}, 909 (1992).

\bibitem{PHH05} S. Pradhan, A.\ Hansen and P.\ C.\ Hemmer, Phys.
Rev. Lett. \textbf{95} 125501 (2005); Phys. Rev. E \textbf{74} 016122 (2006).


\bibitem{HHP} See P.\ C.\ Hemmer, A.\ Hansen, and S.\ Pradhan,
p.\ 27 in Ref. [10]. 

\bibitem{Kun06} F. Raischel, F. Kun and H. J. Herrmann, Phys. Rev. E \textbf{74} 035104 (2006).

\bibitem{Peirce} F.\ T.\ Peirce, J.\ Text.\ Ind. \textbf{17},
355 (1926).

\bibitem{Daniels} H.\ E.\ Daniels, Proc.\ Roy.\ Soc.\ London
\textbf{A 183}, 405 (1945).

\bibitem{Smith} R.\ L.\ Smith, Ann.\ Prob.\ \textbf{10}, 137
(1982).

\bibitem{Phoenix} S.\ L.\ Phoenix and R.\ L.\ Smith, Int.\ J.\ Solids
Struct. \textbf{19}, 479 (1983). 
\end{thebibliography}
\end{document}